\documentclass[aps,prl,final,twocolumn,superscriptaddress,floatfix,preprintnumbers]{revtex4}
\usepackage{dcolumn} 
\usepackage{longtable}%
\usepackage[latin1]{inputenc}
\usepackage{bm}%
\usepackage{docs}
\usepackage{graphics}
\usepackage{epsf}
\usepackage{epsfig}
\usepackage{amssymb}
\usepackage{amsbsy}
\usepackage{amsfonts}
\usepackage[tbtags]{amsmath}

\usepackage{latexsym}
\usepackage[mathscr]{eucal}
\usepackage{graphicx}
\newcommand{\bc}{\begin{center}}
\newcommand{\ec}{\end{center}}
\newcommand{\be}{\begin{equation}}
\newcommand{\ee}{\end{equation}}
\newcommand{\bea}{\begin{eqnarray}}
\newcommand{\eea}{\end{eqnarray}}

\newcommand{\eps}{\epsilon}

\def\Nf{n_f}

\def\t0{t\text{{\ttfamily =}}0}

\begin{document}  

\title{Transverse spin structure of the nucleon from lattice QCD simulations}

\preprint{DESY 06-245, Edinburgh 2006/41, TUM/T39-06-16
}

\author{ M.~G\"ockeler} 
  \affiliation{Institut f\"ur Theoretische Physik, Universit\"at Regensburg, 93040 Regensburg, Germany}
\author{Ph.~H\"agler}
  \affiliation{Institut f\"ur Theoretische Physik T39, Physik-Department der TU M\"unchen, James-Franck-Stra\ss{}e, 85747 Garching, Germany}
   \email{phaegler@ph.tum.de}
\author{R.~Horsley}
  \affiliation{School of Physics, University of Edinburgh, Edinburgh EH9 3JZ, UK}
\author{Y.~Nakamura}
  \affiliation{John von Neumann-Institut f\"ur Computing NIC / DESY, 15738 Zeuthen, Germany}
\author{D.~Pleiter}
  \affiliation{John von Neumann-Institut f\"ur Computing NIC / DESY, 15738 Zeuthen, Germany} 
\author{P.E.L. Rakow} 
\affiliation{Theoretical Physics Division, Department of Mathematical Sciences, University of Liverpool, Liverpool L69 3BX, UK}
\author{A. Sch\"afer} 
  \affiliation{Institut f\"ur Theoretische Physik, Universit\"at Regensburg, 93040 Regensburg, Germany}
\author{G.~Schierholz}
  \affiliation{Deutsches Elektron-Synchrotron DESY, 22603 Hamburg, Germany}
  \affiliation{John von Neumann-Institut f\"ur Computing NIC / DESY, 15738 Zeuthen, Germany}
\author{H.~St\"uben} 
  \affiliation{Konrad-Zuse-Zentrum f\"ur Informationstechnik Berlin, 14195 Berlin, Germany}
 \author{J.M. Zanotti} 
  \affiliation{School of Physics, University of Edinburgh, Edinburgh EH9 3JZ, UK}
  \collaboration{QCDSF/UKQCD Collaborations}

   \date{\today}

\begin{abstract} 
We present the first calculation in lattice QCD of the lowest two moments 
of transverse spin densities of quarks in the nucleon. 
They encode correlations between quark spin and orbital angular momentum. 
Our dynamical simulations are based on two flavors of clover-improved Wilson 
fermions and Wilson gluons.
We find significant contributions from certain quark helicity flip generalized parton distributions,
leading to strongly distorted densities of transversely polarized quarks in the nucleon. 
In particular, based on our results and recent arguments by Burkardt [Phys. Rev. D 72 (2005) 094020], 
we predict that the Boer-Mulders-function $h_1^\perp$, describing correlations of transverse quark spin and
intrinsic transverse momentum of quarks, is large and negative for both up and down quarks.
\end{abstract}

\maketitle

{\em Introduction.}---The transverse spin (transversity)
%
%
structure of the nucleon received a lot of attention in recent years from both theory and 
experiment 
as it provides a new perspective on hadron structure and QCD evolution
(for a review see \cite{Barone:2001sp}). 
A central object of interest is the quark transversity distribution $\delta q(x)=h_1(x)$, 
which describes the probability to find a transversely polarized quark with longitudinal momentum fraction $x$ 
in a transversely polarized nucleon \cite{Jaffe:1991kp}.
Much progress has been made in the understanding of so-called transverse momentum
dependent PDFs (tmdPDFs) like  e.g. the Sivers function $f_{1T}^\perp(x,k_\perp^2)$ \cite{Sivers:1989cc}, 
which measures the correlation
of the intrinsic quark transverse momentum $k_\perp$ and the transverse nucleon spin $S_\perp$, as well as the
Boer-Mulders function $h_1^\perp(x,k_\perp^2)$ \cite{Boer:1997nt}, describing the correlation of $k_\perp$ and 
the transverse quark spin $s_\perp$. 
While the Sivers function begins to be understood, still
very little is known about the sign and size of the Boer-Mulders function.

A particularly promising approach is based on 3-dimensional densities of quarks in the nucleon, 
$\rho(x,b_\perp,s_\perp,S_\perp)$ \cite{Diehl:2005jf},
representing the probability to find a quark with momentum fraction $x$ and
transverse spin $s_\perp$ at distance $b_\perp$ from the center-of-momentum of the nucleon with transverse
spin $S_\perp$.
As we will see below, these transverse spin densities show
intriguing correlations of transverse coordinate and spin degrees of freedom.
According to Burkardt \cite{Burkardt:2003uw,Burkardt:2005hp}, they
are directly related to the above mentioned Sivers- and Boer-Mulders-functions. 
Our lattice results on transverse spin densities therefore provide
for the first time 
quantitative predictions
for the signs and sizes of these tmdPDFs and the corresponding experimentally accessible asymmetries.

Lattice calculations give access to $x$-moments of
transverse quark spin densities
\cite{Diehl:2005jf}
\begin{align}
\rho^{n}&(b_\perp,s_\perp,S_\perp)=\int_{-1}^{1}dx\,x^{n-1}\rho(x,b_\perp,s_\perp,S_\perp) 
= \nonumber \\
 & \frac{1}{2}\left\{A_{n0}(b_\perp^2)
+ s_\perp^i S_\perp^i \left( A_{Tn0}(b_\perp^2)
- \frac{1}{4m^2} \Delta_{b_\perp} \widetilde{A}_{Tn0}(b_\perp^2) \right) \right.  \nonumber \\
&+ \frac{ b_\perp^j \eps ^{ji}}{m} \left( S_\perp^i B_{n0}'(b_\perp^2)
+ s_\perp^i \overline{B}_{Tn0}'(b_\perp^2) \right) \nonumber \\
&+ \left. s_\perp^i ( 2 b_\perp^i b_\perp^j
- b_\perp^2 \delta^{ij} ) S_\perp^j \frac{1}{m^2} \widetilde{A}_{Tn0}''(b_\perp^2)
\right\}, \hspace{-3mm} \label{density}
\end{align}
where $m$ is the nucleon mass.
The $b_\perp$-dependent nucleon generalized form factors (GFFs) $A_{n0}(b_\perp^2)$, $A_{Tn0}(b_\perp^2),\ldots$  in Eq.~(\ref{density}) are related to GFFs in momentum space $A_{n0}(t)$, $A_{Tn0}(t),\ldots$
by a Fourier transformation 
\begin{eqnarray}
\vspace{-1mm}
f(b_\perp^2)
\equiv\int \frac{d^2\Delta_\perp}{(2\pi)^2} e^{-i b_\perp \cdot \Delta_\perp} f(t=-\Delta_\perp^2)\,,
\label{Fourier}
\vspace{-1mm}
\end{eqnarray}
where
$\Delta_\perp$ is the transverse momentum transfer to the nucleon.
Their derivatives 
are defined by $f' \equiv \partial_{b_\perp^2}f$ and
$\Delta_{b_\perp}f \equiv 4\partial_{b_\perp^2}\big(b_\perp^2\partial_{b_\perp^2}\big)f$.
The generalized form factors in this work are directly related to $x$-moments of the corresponding 
vector and tensor generalized parton distributions (GPDs) (for a review see \cite{Diehl:2003ny}). 
The probability interpretation of GPDs 
in impact parameter space has been first noted in \cite{Burkardt:2000za}. 
Apart from the orbitally symmetric \textsl{monopole} terms in the second line of Eq.(\ref{density}),
there are two \textsl{dipole} structures present in the third line of Eq.(\ref{density}),
$b_\perp^j \eps^{ji} s_\perp^i$ and $b_\perp^j \eps^{ji} S_\perp^i$.
The fourth line in Eq.(\ref{density}) corresponds to a \textsl{quadrupole} term. The (derivatives of the) three
GFFs $B_{n0}(b_\perp)$,  $\overline B_{Tn0}(b_\perp)$ and $\widetilde A_{Tn0}(b_\perp)$ 
thus determine how strongly the orbital symmetry in the transverse plane 
is distorted by the dipole and the quadrupole terms. 

The GFFs $A_{n0}(t)$, $A_{Tn0}(t),\ldots$ 
parametrize off-forward nucleon matrix elements of certain local quark operators.
For the lowest moment $n=1$ one finds $A_{10}(t)=F_1(t)$, $B_{10}(t)=F_2(t)$ and $A_{T10}(t)=g_T(t)$ where 
$F_1$, $F_2$ and $g_T$ are the Dirac, Pauli and tensor nucleon form factors, respectively.
A concrete example of the corresponding parametrization for $n=1$ is given by \cite{Diehl:2001pm,Hagler:2004ytChen:2004cg}
\begin{equation}
\begin{split}
\left\langle P^{\prime}\Lambda^{\prime}\right|&
 {\mathcal O}_T^{\mu\nu} \left| P \Lambda\right\rangle
=\overline u(P',\Lambda ') \bigg\{ \sigma^{\mu\nu}\gamma_5
   \bigg(\!\! A_{T10}(t) \\
   &- \frac{t}{2m^2}\widetilde A_{T10}(t) \!\!\bigg)
 + \frac{\eps^{\mu\nu\alpha\beta} \Delta_{\alpha} \gamma_{\beta}} {2 m} \overline B_{T10}(t) \\
 &- \frac{\Delta^{[\mu} \sigma^{\nu]\alpha}\gamma_5 \Delta_{\alpha}} {2m^2 } \widetilde A_{T10}(t)\!\bigg\}
 u(P,\Lambda)\ ,
 \end{split}
\label{tn0}
\end{equation}
where ${\mathcal O}_T^{\mu\nu}=\bar{q}\sigma^{\mu \nu }\gamma_5 q$ 
is the lowest element of the tower of local leading twist tensor 
(quark helicity flip) operators.
Parametrizations for higher moments  $n\ge 1$ in terms of tensor GFFs and their relation to GPDs are given in
\cite{Hagler:2004ytChen:2004cg}.
%
%
As it is very challenging to 
access
tensor GPDs in 
experiment 
\cite{Collins:1999un},
input from lattice QCD calculations is crucial in this case.

%
\begin{figure}[t]
\vspace{-1mm}
\bc
\includegraphics[width=7.cm,angle=0,clip=true]{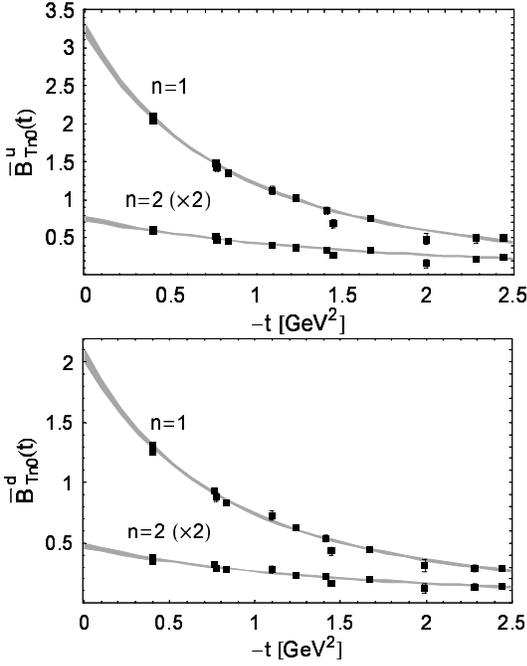}
\caption{Results for the generalized form factors $\overline B_{T(n=1,2)0}(t)$. The
corresponding p-pole parametrizations are shown by the shaded error bands.} 
\label{FigBbarT10}
\ec
\vspace{-4mm}
\end{figure}
%
%
{\em Simulation results.}---Our lattice calculations are based on configurations generated with 
$\Nf=2$ 
dynamical non-perturbatively ${\mathcal O}(a)$ improved Wilson fermions and Wilson gluons.
Simulations have been performed at four different couplings
$\beta=5.20$, $5.25$, $5.29$, $5.40$ with up to five different $\kappa=\kappa_{\mathrm {sea}}$ values per $\beta$,
on lattices of $V\times T=16^3\times 32$ and $24^3\times 48$.
The lattice spacings are below $0.1$~fm, the range of pion masses extends down to ~$400$~MeV
and the spatial volumes are as large as ~$(2.1 \text{ fm})^3$.
The lattice scale $a$ in physical units has been set 
using a Sommer scale of $r_0=0.467$~fm \cite{Khan:2006de,Aubin:2004wf}.
The computationally demanding disconnected contributions are not included. 
We expect,
however, that they are small for the tensor GFFs \cite{Gockeler:2005cj}.
We use non-perturbative renormalization \cite{reno} to transform the lattice results to the 
$\overline{\mbox{MS}}$ scheme at a scale of $4$~GeV$^2$.
The calculation of GFFs in lattice QCD follows standard 
methods
(see, e.g., \cite{Gockeler:2003jf,MIT,Gockeler:2005aw}). 

In Fig.~\ref{FigBbarT10}, we show 
as an example
results for the GFFs $\overline B^{u,d}_{T(n=1,2)0}(t)$, 
corresponding to the lowest two moments $n=1,2$ of the GPD $\overline E^{u,d}_T(x,\xi,t)$ \cite{Diehl:2005ev},
as a function of the momentum transfer squared $t$, for a pion mass of $m_\pi\approx 600$~MeV,
a lattice spacing of $a\approx 0.08$~fm and a volume of $V\approx (2$~fm$)^3$.
For the extrapolation to the forward limit ($t=0$) and in order to get a functional 
parametrization of the lattice results, we fit all GFFs 
using a p-pole ansatz
$F(t)=F_0/( 1 - (t/(p\,m_p^2))^p$
%
%
with the three parameters  $F_0=F(\t0)$, $m_p$ and $p$ for each GFF.
We consider this ansatz \cite{Brommel:2006ww} to be more physical than
previous ones as the rms-radius $\langle r^2 \rangle^{1/2}\propto m_p^{-1}$ is independent of $p$.
%
\begin{figure}[t]
\vspace{-1mm}
\bc
\includegraphics[width=8.cm,angle=0]{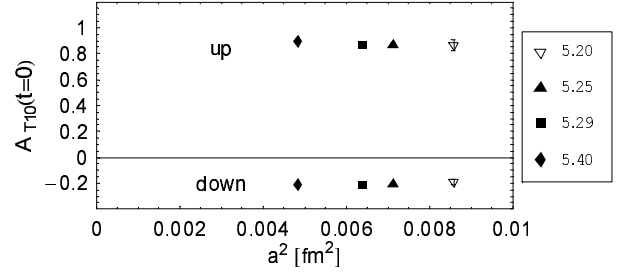}
\caption{Study of discretization errors of the tensor charge $A_{T10}(\t0)=g_T(\t0)$ for
up- and down-quarks at a pion mass of $m_\pi\approx 600$~MeV.}
\label{FigDiscrErr}
\ec
\vspace{-4mm}
\end{figure}
It turns out that in most cases the statistics is not sufficient to determine
all three parameters from a single fit to the lattice data. For a given generalized form factor,
we therefore fix the power $p$ first, guided by fits to selected datasets, and subsequently
determine the forward value $F_0$ and the p-pole mass $m_p$ by a full fit to the lattice data.
Some GFFs show a quark flavor dependence of the value of $p$, which has already
been observed in \cite{Pleiter:2006Lat} for the Dirac form factor. 
\begin{figure}[t]
\vspace{-1mm}
\bc
\includegraphics[width=6.5cm,angle=0]{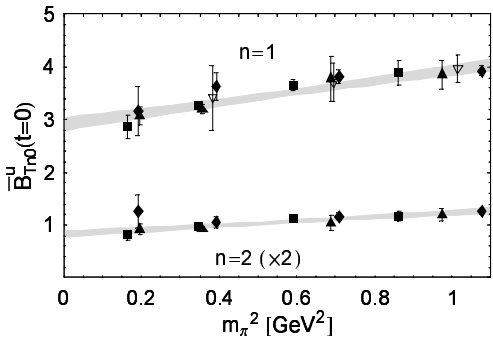}
\caption{Pion mass dependence of the generalized form factors $\overline B_{T(n=1,2)0}(\t0)$
for up-quarks. The shaded error bands show extrapolations
to the physical pion mass based on an ansatz linear in $m_\pi^2$. The symbols are as in Fig.~2.}
\label{FigBbarT10mPi2}
\ec
\vspace{-4mm}
\end{figure}
For the examples in Fig.~\ref{FigBbarT10}, we find for u-quarks $\overline B^u_{T10}(\t0)=3.34(8)$ with
$m_p=0.907(75)$ ~GeV, $\overline B^u_{T20}(\t0)=0.750(32)$ with
$m_p=1.261(40)$ ~GeV and for d-quarks $\overline B^d_{T10}(\t0)=2.06(6)$ with $m_p=0.889(48)$ ~GeV,
$\overline B^d_{T20}(\t0)=0.473(22)$ with $m_p=1.233(27)$ ~GeV (all for $p=2.5$).
We have checked that the final p-pole parametrizations only show a mild dependence on 
the value of $p$
chosen prior to the fit.
In order to see to what extent our calculation 
is affected by discretization errors, we plot as an example in Fig.~\ref{FigDiscrErr} the tensor charge $A_{T10}(\t0)=g_T(\t0)$ versus the lattice spacing squared, for a fixed $m_\pi\approx600$~MeV.
The discretization errors seem to be smaller than the statistical errors,
and we will neglect any dependence of the GFFs on $a$ in the following.
Taking our investigations of the volume dependence of the nucleon mass
and the axial vector form factor $g_A$ \cite{Khan:2006de,AliKhan:2003cu} as a guide, 
we estimate that the finite volume effects for the lattices and observables 
studied in this work are small and may be neglected. 
%

As an example of the pion mass dependence of our results,
we show in Fig.~\ref{FigBbarT10mPi2} the GFFs $\overline B^u_{T(n=1,2)0}(\t0)$
versus $m_\pi^2$.  
Unfortunately we cannot expect chiral perturbation theory predictions \cite{Ando:2006skDiehl:2006ya}
to be applicable to most of our lattice data points, for which the pion mass is still rather large.
To get an estimate of the GFFs at the physical point, we extrapolate the forward moments and
the p-pole masses using an ansatz linear in $m_\pi^2$. 
The results of the corresponding fits 
are shown as shaded error bands in Fig.~\ref{FigBbarT10mPi2}. 
At $m_\pi^\text{phys}=140$~MeV, we find $\overline B^u_{T10}(\t0)=2.93(13)$, $\overline B^d_{T10}(\t0)=1.90(9)$
and $\overline B^u_{T20}(\t0)=0.420(31)$, $\overline B^d_{T20}(\t0)=0.260(23)$.
These comparatively large values already indicate a significant 
impact of this tensor
GFF on the transverse spin structure of the nucleon, as will be discussed below.
Since the (tensor) GPD $\overline E_T$ can be seen as the analogue of the
(vector) GPD $E$, we may define an anomalous tensor magnetic moment \cite{Burkardt:2005hp},
$\kappa_T\equiv \int dx \overline E_T(x,\xi,\t0)=\overline B_{T10}(\t0)$, similar to the standard anomalous magnetic moment $\kappa= \int dx E(x,\xi,\t0)= B_{10}(\t0)=F_2(\t0)$. 
While the u- and d-quark contributions to the anomalous magnetic moment are both large and of opposite
sign, $\kappa^\text{up}_\text{exp}\approx1.67$
and $\kappa^\text{down}_\text{exp}\approx-2.03$, we find large positive values for the
anomalous tensor magnetic moment for both flavors, $\kappa^\text{up}_\text{T,latt}\approx3.0$ and $\kappa^\text{down}_\text{T,latt}\approx1.9$.
Similarly large positive values have been obtained in a recent model calculation \cite{Pasquini:2005dk}. 
Large $N_c$ considerations predict $\kappa^\text{up}_\text{T}\approx \kappa^\text{down}_\text{T}$ \cite{Burkardt:2006td}.
%
%
\begin{figure}[t]
\vspace{+1mm}
\bc
\includegraphics[width=8.5cm,angle=0]{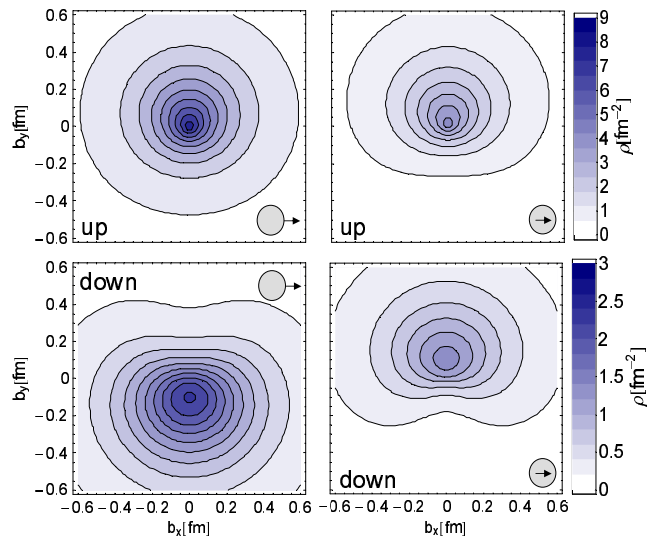}
\caption{Lowest moment ($n=1$) of the densities of unpolarized quarks in a transversely polarized nucleon (left) and transversely polarized quarks in an unpolarized nucleon (right) for up (upper plots) and down (lower plots) quarks.
The quark spins (inner arrows) and nucleon spins (outer arrows) are oriented in the transverse plane as indicated.}
\label{densities1}
\ec
\vspace{-5mm}
\end{figure}

Let us now discuss our results for $\rho^{n}(b_\perp,s_\perp,S_\perp)$ in Eq.~(\ref{density}).
For the numerical evaluation 
we Fourier transform the p-pole
parametrization 
to impact parameter ($b_\perp$) space.
The parametrizations of the impact parameter dependent
GFFs
then depend only on the p-pole masses $m_p$ and the forward values $F_0$.
Before showing our final results, 
we would like to note that the moments
of the transverse spin density can be written as sum/difference of the corresponding 
moments for
quarks and antiquarks, $\rho^n=\rho_q^n+(-1)^n \rho_{\overline q}^n$, because vector 
and tensor
operators transform identically under charge conjugation. 
Although we expect contributions from
antiquarks to be small in general, only the $n$-even moments must be strictly positive.
In Fig.~\ref{densities1}, we show the lowest moment $n=1$ of spin densities for up and down
quarks in the nucleon. Due to the large anomalous magnetic moments $\kappa^{u,d}$, 
we find strong distortions for unpolarized quarks in transversely polarized nucleons (left part of the figure).
This has already been discussed in \cite{Burkardt:2003uw}, and can serve as a dynamical explanation of the experimentally observed
Sivers-effect.
\begin{figure}[t]
\vspace{-1mm}
\bc
\includegraphics[width=8.5cm,angle=0]{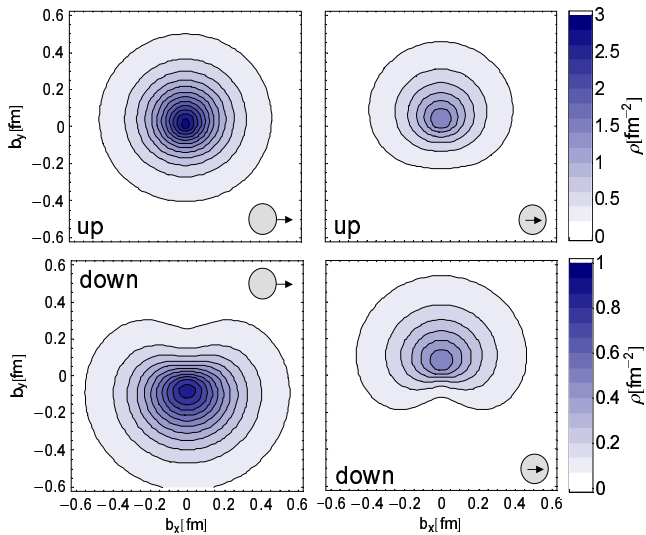}
\caption{Second moment ($n=2$) of transverse spin densities. For details see caption of Fig.~4.}
\label{densities2}
\ec
\vspace{-4mm}
\end{figure}
Remarkably, we find even stronger distortions for
transversely polarized quarks $s_\perp=(s_x,0)$ in an unpolarized nucleon, as can be seen on
the right hand side of
Fig.~\ref{densities1}. 
The densities for up and for down quarks in this case 
are both deformed in positive $b_y$ direction due to the large positive values for the tensor 
GFFs $\overline B^u_{T10}(\t0)$ and $\overline B^d_{T10}(\t0)$, in strong contrast to 
the distortions one finds for unpolarized quarks in a transversely polarized nucleon.
All of these observations are actually quite plausible, because
there is no gluon transversity which could mix with quarks under evolution.
Therefore, the transverse spin structure is much more valence-like than the 
longitudinal one, which is strongly affected by the negative sign of the 
photon-gluon contribution. Thus the transverse quark spin and the transverse quark orbital
angular momentum simply seem to be aligned. 
The fact that the u-quark (d-quark) spin is
predominantly oriented parallel (antiparallel) to the nucleon spin
then explains why densities of quarks with spin in $x$-direction in an unpolarized
nucleon (moving towards the observer in $z$-direction) are larger in the upper half plane.
However, the contributions from down quarks with spin in $(-x)$-direction dominate in 
a polarized nucleon with spin along the $(+x)$-axis, 
such that the orbital motion around the $(-x)$-direction
leads to a larger down quark density in the lower half plane.
It has been argued by Burkardt \cite{Burkardt:2005hp} that the deformed densities
on the right hand side of Fig.~\ref{densities1} are related to a non-vanishing
Boer-Mulders function \cite{Boer:1997nt} $h_1^{\perp}$ which describes the correlation
of intrinsic quark transverse momentum and the transverse quark spin $s_\perp$.
According to \cite{Burkardt:2005hp} we have in particular $\kappa_T \sim - h_1^{\perp}$. 
If this conjecture is correct
our results imply that
the Boer-Mulders function is large and negative
both for up and down quarks. 
The fact that the correlation of quark and nucleon spin is not 100 percent explains
why the deformation is more pronounced in the Boer-Mulders than in the Sivers case.

Fig.~\ref{densities2} shows 
the $n=2$-moment of the densities.
Obviously, the pattern is very similar to that in Fig.~\ref{densities1}, which supports our simple interpretation. 
The main difference is that the
densities for the higher $n=2$-moment are more peaked around the origin $b_\perp=0$
as already observed in \cite{MIT-2} for the vector and axial vector GFFs.
%

{\em Conclusions.}---We have presented first lattice results for the lowest two moments of 
transverse spin densities of quarks in the nucleon.
Due to the large and positive contributions from the tensor GFF $\overline B_{Tn0}$ 
for up and for down quarks, we find strongly distorted spin densities
for transversely polarized quarks in an unpolarized nucleon. According 
to Burkardt \cite{Burkardt:2005hp}, this leads to the prediction of a sizable
negative Boer-Mulders function \cite{Boer:1997nt} for up and down quarks,
which may be confirmed in experiments at e.g. JLab and GSI/FAIR 
\cite{Avakian2006,Kotulla:2004ii}.
\begin{acknowledgments}
The numerical calculations have been performed on the Hitachi SR8000 at LRZ (Munich),
the apeNEXT at NIC/DESY (Zeuthen) and the
BlueGene/L at NIC/FZJ (Jülich), EPCC (Edinburgh) and 
KEK (by the Kanazawa group as part of the DIK research programme).
This work was supported by DFG (Forschergruppe
Gitter-Hadronen-Ph\"anomenologie and Emmy-Noether programme), 
HGF (contract No. VH-NG-004) and EU I3HP (contract No. RII3-CT-2004-506078).
\vspace{-1mm}
\end{acknowledgments}
\vspace{-3mm}

\end{document}